\documentclass[
 reprint,amsmath,amssymb,aps,
prl,
]{revtex4-2}

\usepackage{graphicx}
\usepackage{dcolumn}
\usepackage{bm}
\usepackage{hyperref}
\hypersetup{colorlinks, linkcolor={blue}, citecolor={blue}, urlcolor={blue}}
\usepackage{xcolor}
\usepackage{pifont}
\newcommand{\cM}{\boldsymbol{\mathcal M}}
\newcommand{\cN}{\boldsymbol{\mathcal N}}
\newcommand{\btau}{\boldsymbol{\tau}}

\begin{document}

\title{Pseudospin Dynamics of Charge Order}
\author{Ping Tang}
\email{tang.ping.a2@tohoku.ac.jp}
\affiliation{Institute for Materials Research, Tohoku University, Sendai 980-8577, Japan}
\date{\today}

\begin{abstract}
Charge order is conventionally viewed as a static modulation of the electronic density, despite growing experimental capabilities to probe and manipulate its nonequilibrium evolution. In contrast to spin-ordered systems, a microscopic framework for charge-order dynamics and its control under external driving remains largely underdeveloped. Here, starting from an extended Hubbard model, we derive an effective pseudospin model in which the charge-ordered state maps onto staggered pseudospin order. The resulting charge-order dynamics is governed by Landau--Lifshitz--Gilbert-like equations for the pseudospins, closely analogous to those of a bipartite antiferromagnet. We show that an external electric field directly controls the staggered pseudospin order and, above a threshold field, drives coherent reversal of the charge-order polarity by destabilizing collective pseudospin excitations. Our results establish the charge pseudospin as a microscopic dynamical degree of freedom for coherent switching and control of charge order, providing a charge-sector analogue of the well-established framework for spin-order dynamics in spintronics.
\end{abstract}
\maketitle

\emph{Introduction.---}Charge order is a ubiquitous broken-symmetry state in strongly correlated
electron systems~\cite{RevModPhys.70.1039,RevModPhys.87.457}, which occurs in transition-metal
oxides~\cite{PhysRevLett.82.4679,tokura2000orbital,Attfield2006,comin2015symmetry,comin2016resonant}, organic
conductors~\cite{PhysRevB.62.R7679,dressel2004optical,takahashi2006charge}, mixed-valence
compounds~\cite{PhysRevB.62.12167,renner2002atomic,PhysRevB.81.134417,Senn2012}, and low-dimensional quantum
materials~\cite{Seo2004,flicker2015charge,hwang2024charge}. A paradigmatic description is
provided by the extended Hubbard model~\cite{PhysRevLett.53.2327,PhysRevB.39.9397,PhysRevLett.106.236805,PhysRevLett.110.166401,PhysRevLett.111.036601,PhysRevB.95.115149,Paki2019}, in
which the competition between the on-site ($U$) and intersite ($V$) Coulomb
interactions gives rise to a rich phase diagram encompassing antiferromagnetic
and charge-ordered phases \cite{PhysRevLett.53.2327,PhysRevB.39.9397,PhysRevB.95.115149,Paki2019}. Extensive theoretical and experimental efforts have established the equilibrium physics of charge-ordered states, including their phase diagrams and intricate interplay with magnetic, orbital, and superconducting orders~\cite{Attfield2006,Oles2010,Comin2016,Frano2020}. Nonequilibrium studies, on the other hand, have focused primarily on depinning and sliding~\cite{PhysRevB.19.3970,Gruner1988,Monceau2012}, ultrafast melting~\cite{fiebig2000sub,Iwai2007,PhysRevLett.103.155702,PhysRevLett.105.187401}, and transitions between charge-ordered and other electronic phases~\cite{Stojchevska2014,Geremew2019}. 

By contrast, the dynamics and manipulation of magnetic or spin order constitute a central theme of spintronics~\cite{Fert2008,RevModPhys.76.323}. The associated
magnetization dynamics is governed by the Landau--Lifshitz--Gilbert equation,
which captures the precession and damping of the magnetic order parameter in an effective
magnetic field~\cite{Landau1935,Gilbert2004}. Magnetic order can be switched
by external magnetic fields or electrically through spin-transfer torques
exerted by spin-polarized conduction
electrons~\cite{Slonczewski1996,Berger1996,Ralph2008}. Very recently, Kikkawa \textit{et al.}~\cite{kikkawa2026electronic} reported electrical reversal of the charge-order polarity in semiconducting LuFe$_2$O$_4$~\cite{Ikeda2005}, offering a charge-sector analogue of electric-current-induced magnetization switching~\cite{Katine2000,Miron2011Nature,Liu2012Science}. Despite this phenomenological analogy, a microscopic connection to magnetization switching is far from straightforward, since the underlying order parameters are fundamentally different. Magnetic order is vectorial and breaks time-reversal symmetry, whereas charge order is characterized by a scalar modulation or disproportionation of the electronic density that breaks translational or intracell symmetry. Whether charge order nevertheless admits a collective dynamical description analogous to spin dynamics, and how its coherent polarity reversal under external driving can be described microscopically, therefore remain open questions.

In this Letter, we address these questions by deriving the collective
pseudospin dynamics of charge order from the extended Hubbard model.
We show that the low-energy charge sector maps onto an effective
pseudospin Hamiltonian in which the charge-ordered state corresponds
to staggered pseudospin order. The resulting charge-order dynamics
is governed by Landau--Lifshitz--Gilbert-like equations for the
pseudospins, closely analogous to those of a bipartite
antiferromagnet. An external electric field acts as a staggered field
on the pseudospins and, above a threshold value, enables coherent
reversal of the charge-order polarity by triggering a dynamical
instability of the collective pseudospin excitations. Our results establish
a microscopic framework for the collective dynamics and electrical
switching of the charge order, extending concepts of
coherent spin dynamics and control to the charge sector.

\begin{figure}
    \centering
    \includegraphics[width=8.6 cm]{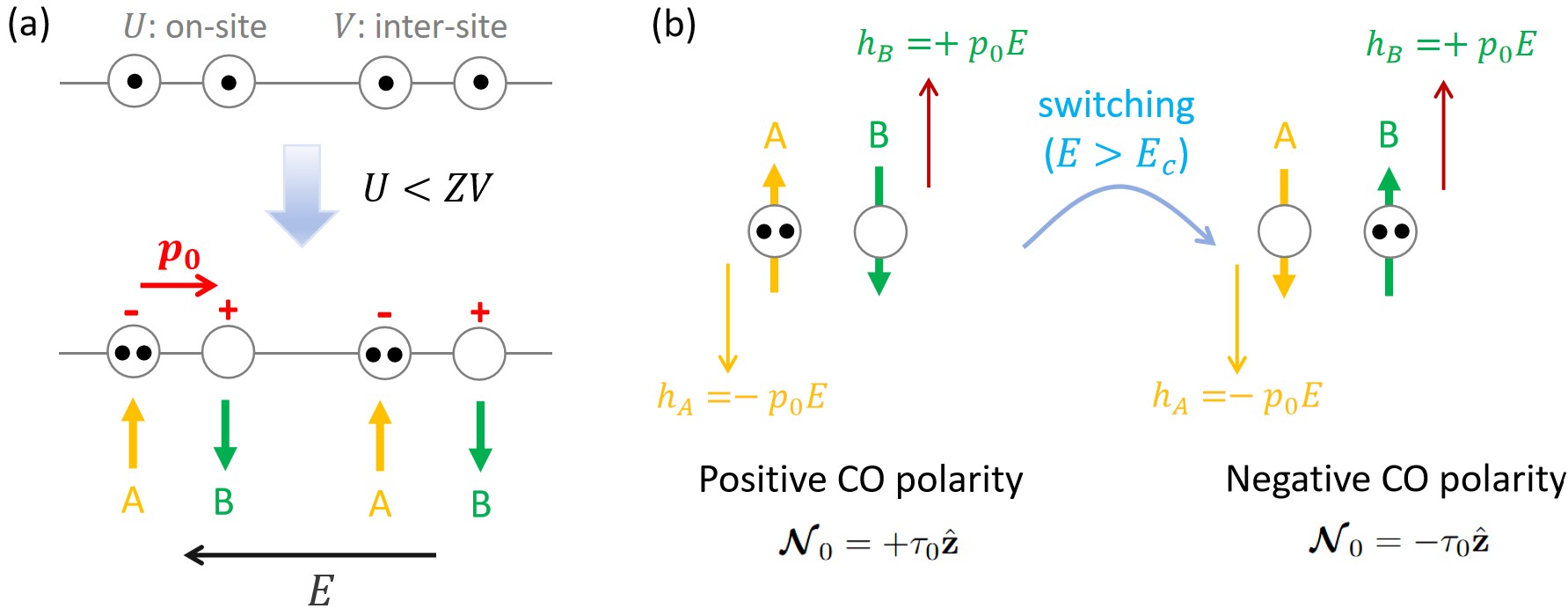}
    \caption{(a) Charge order (CO) with alternating doubly occupied and empty sites in a half-filled extended Hubbard model with $U<ZV$, in which an intracell electric
dipole $p_0$ emerges due to the accompanying inversion-symmetry breaking. (b) Switching of the CO polarity by an external electric field that couples
to the intracell dipoles and thereby acts as a staggered field on the two
pseudospin sublattices. Here, doubly occupied (empty) sites are represented
by orange (green) pseudospin arrows, and the CO state corresponds to
staggered pseudospin order on the $A$ and $B$ sublattices. }
    \label{Fig-CO}
\end{figure}

\emph{Pseudospin model.---}We begin with the extended Hubbard model for describing charge ordering in correlated electron systems~\cite{PhysRevLett.53.2327,PhysRevB.39.9397, PhysRevLett.106.236805,PhysRevLett.110.166401,PhysRevLett.111.036601,PhysRevB.95.115149,Paki2019},
\begin{align}
H=&-t\sum_{\langle ij\rangle,\sigma}\left(c_{i\sigma}^{\dagger}c_{j\sigma}+\mathrm{H.c.}\right)
-\mu\sum_i n_i \nonumber\\&+U\sum_i n_{i\uparrow}n_{i\downarrow}+V\sum_{\langle ij\rangle}n_i n_j,
\label{eq:Hubbard}
\end{align}
where $c_{i\sigma}^{(\dagger)}$ annihilates (creates) an electron with spin $\sigma$ at site $i$, $n_{i\sigma}=c_{i\sigma}^{\dagger}c_{i\sigma}$, and $n_i=n_{i\uparrow}+n_{i\downarrow}$. Here, $t$ denotes the nearest-neighbor hopping amplitude, $\mu$ is the chemical potential, and $U$ and $V$ are the on-site and nearest-neighbor Coulomb repulsions, respectively. In the conventional Hubbard model at half filling, a sufficiently strong on-site repulsion $U\gg t$ suppresses double occupancy and favors a Mott-insulating state with antiferromagnetic spin correlations. The additional intersite repulsion $V$, by contrast, favors charge disproportionation between neighboring sites, leading to competition between antiferromagnetically correlated Mott and charge-ordered states depending on the relative strengths of $U$ and $V$. In the atomic limit $t_{ij}=0$, the singly occupied Mott state has an energy per site $E_{\rm AFM}=ZV/2-\mu$, whereas the charge-ordered state with alternating empty and doubly occupied sites has $E_{\rm CO}=U/2-\mu$, where $Z$ is the lattice coordination number; the latter is therefore energetically favored for $U<ZV$. The extended Hubbard model has consequently been widely employed as a minimal model of correlation-driven charge order in strongly correlated electron systems~\cite{PhysRevB.49.9670,PhysRevB.95.115149,PhysRevLett.53.2327,Seo2000, PhysRevB.70.235107}.

We focus on the strong charge-ordering regime near half filling, in which the relevant low-energy local configurations are the empty state $|0\rangle_i$ and the doubly occupied state $|2\rangle_i=c_{i\uparrow}^{\dagger}c_{i\downarrow}^{\dagger}|0\rangle_i$ that alternate between neighboring sites, whereas singly occupied states occur only as virtual intermediate states. Within this projected subspace, we introduce charge-pseudospin-$1/2$ operators
\begin{equation}
\tau_i^z=\frac{n_i-1}{2},\qquad
\tau_i^+= c_{i\uparrow}^{\dagger}c_{i\downarrow}^{\dagger},\qquad
\tau_i^-=c_{i\downarrow}c_{i\uparrow},
\label{eq:pseudospin_def}
\end{equation}
such that $|0\rangle_i$ and $|2\rangle_i$ are eigenstates of $\tau_i^z$ with eigenvalues $-1/2$ and $+1/2$, respectively, and are connected by the pseudospin ladder operators through $\vert0\rangle_{i}=\tau_{i}^{-}|2\rangle_i$ and $\vert 2\rangle_{i}=\tau_{i}^{\dagger}|0\rangle_i$. The longitudinal component $\tau_i^z$ characterizes the local charge configuration, whereas the transverse components, or equivalently $\tau_i^\pm$, describe quantum coherence and transitions between the empty and doubly occupied states. Within the charge-order subspace ${|0\rangle_i,|2\rangle_i}$, where $n_{i\uparrow}n_{i\downarrow}
=
\tau_i^z+1/2$, the non-hopping part $H_0$ of the Hamiltonian reduces to
\begin{equation}
PH_{0}P=4V\sum_{\langle ij \rangle}\tau_{i}^{z}\tau_{j}^{z}-h_{z}\sum_{i}\tau_{i}^{z} +\text{const.}  \label{H0}
\end{equation}
where $P=\prod_i\left(|0\rangle_i{}_i\langle0|+|2\rangle_i{}_i\langle2|\right)$ is the projector onto the charge-pseudospin subspace, and $h_z=2\mu-U-2ZV$ is an effective longitudinal field conjugate to the pseudospins. At the particle-hole-symmetric half-filled point, $\mu=U/2+ZV$ and hence $h_z=0$, whereas a nonzero $h_z$ corresponds to electron or hole doping away from half filling. The intersite Coulomb repulsion $V>0$ generates an antiferromagnetic Ising coupling in pseudospin space, favoring alternating empty $|0\rangle_i$ and doubly occupied $|2\rangle_i$ sites and thereby stabilizing the charge-ordered state.

The hopping term in Eq.~(\ref{eq:Hubbard}) induces virtual transitions from the charge-ordered states $|2\rangle_i|0\rangle_j$ and $|0\rangle_i|2\rangle_j$ to singly occupied configurations $|1\rangle_i|1\rangle_j$ on neighboring sites $i$ and $j$. Projecting out these high-energy virtual states, the second-order Schrieffer--Wolff transformation~\cite{Schrieffer1966,Bravyi2011} yields an effective pseudospin Hamiltonian of the form~\cite{SM}
\begin{equation}
H_{\text{eff}}=\sum_{\langle ij\rangle}\left[J_z\tau_i^z\tau_j^z-J_\perp\left(\tau_i^x\tau_j^x+\tau_i^y\tau_j^y\right)\right]-h_z\sum_i\tau_i^z
\label{eq:Hps}
\end{equation}
where $J_z\simeq 4(V+t^2/\Delta)$ and $J_\perp\simeq 4t^2/\Delta$, with $\Delta$ denoting the energy required to create the virtual singly occupied configurations~\cite{energy}. The transverse pseudospin components are defined as $\tau_i^x=(\tau_i^++\tau_i^-)/2$ and $\tau_i^y=(\tau_i^+-\tau_i^-)/(2i)$, with the different components obeying the standard angular-momentum commutation relations, $[\tau_{i}^{\alpha},\tau_{j}^{\beta}]=i\delta_{ij}\epsilon_{\alpha\beta\gamma}\tau_{i}^{\gamma}$, where $\epsilon_{\alpha\beta\gamma}$ is the Levi-Civita symbol. The virtual hopping processes generate a ferromagnetic exchange coupling between the transverse pseudospin components, which describes coherent pair transfer between the two charge-ordered configurations $|2\rangle_i|0\rangle_j\leftrightarrow|0\rangle_i|2\rangle_j$, while simultaneously enhancing the antiferromagnetic Ising-like coupling through the second-order energy correction to the charge-ordered configurations. In contrast, in the conventional Hubbard model, the hopping induces isotropic antiferromagnetic correlations between physical spins in the singly occupied manifold~\cite{Takahashi1977,MacDonald1988}. The opposite sign of the transverse pseudospin exchange follows from the fact that the electronic wavefunction is symmetric under the coherent pair transition $\vert 0\rangle_i\vert 2\rangle_j\rightarrow\vert 2\rangle_i\vert 0\rangle_j$, whereas it is antisymmetric under the exchange of different singly occupied states, $\vert \uparrow\rangle_i\vert \downarrow\rangle_j\rightarrow\vert \downarrow\rangle_i\vert \uparrow\rangle_j$. For $J_z>J_\perp>0$, Eq.~(\ref{eq:Hps}) favors an easy-axis antiferromagnetic pseudospin ground state, corresponding physically to alternating empty and doubly occupied sites, i.e., staggered $2/0$ charge configurations. While $J_z$ stabilizes the charge-ordered state, $J_\perp$ enables coherent pair transfer between neighboring charge configurations and thereby underlies the collective dynamics of charge order. The longitudinal field $h_z$ controls the average electron density relative to half filling by inducing a finite macroscopic pseudospin polarization associated with doping.

\emph{Semiclassical pseudospin dynamics.---}We now consider the dynamics of the charge-ordered state in the presence of an external electric field. The charge order admits two energetically degenerate ground states characterized by opposite staggered pseudospin order along the $z$ axis. When accompanied by inversion-symmetry breaking, as illustrated in Fig.~\ref{Fig-CO}, the two charge-order states carry opposite electric dipoles arising from the intracell asymmetric charge distribution, as occurs in electronic ferroelectrics~\cite{PhysRevB.54.17452,ishihara2010electronic}. An electric field $E$ applied along the dipole orientation can therefore lift the degeneracy between the two oppositely polarized charge-order configurations through
\begin{equation}
H_{\text{ext}}=-p_{0}E\left(
\sum_{i\in A}\tau_i^z-\sum_{i\in B}\tau_i^z
\right),
\label{eq:HE}
\end{equation}
where $p_{0}$ is the effective electric dipole moment measuring the coupling strength with the charge order~\cite{notedipole}. Here, $A$ and $B$ label the two sublattices of the staggered pseudospin order. In contrast to $h_z$, the electric field acts as a staggered longitudinal field and couples directly to the charge order.

The dynamics of the charge order are encoded in the Heisenberg equations of motion for the pseudospin operators, $i\hbar\dot{\boldsymbol{\tau}}_{i}=[\boldsymbol{\tau}_{i}, H_{\text{eff}}]$. Including phenomenological damping, the mean-field semiclassical dynamics of the two sublattice pseudospins can be written in a Landau--Lifshitz--Gilbert-like form
\begin{equation}
\hbar\langle\dot{\bm\tau}_i\rangle
=
\langle\bm\tau_i\rangle\times\boldsymbol{\mathcal{B}}_i-\frac{\alpha_G\hbar}{\tau_{0}}\langle\boldsymbol{\tau}_{i}\rangle\times\langle\dot{\boldsymbol{\tau}}_{i}\rangle, 
\label{eq:sublatticeEOM}
\end{equation}
where $\langle\boldsymbol{\tau}_i\rangle$ is the expectation value of the pseudospin operator, with $\tau_0=|\langle\boldsymbol{\tau}_i\rangle|$ its length, $\boldsymbol{\mathcal{B}}_i=-\langle\partial H_{\text{eff}}/\partial\boldsymbol{\tau}_{i}\rangle$ the effective pseudospin field acting on $\langle\boldsymbol{\tau}_{i}\rangle$, and $\alpha_G$ the dimensionless Gilbert damping. Within the macro-pseudospin approximation~\cite{Ralph2008}, $\boldsymbol{\mathcal{B}}_{A}=-Z\hat{\mathcal{J}}\cdot\langle \boldsymbol{\tau}_{B}\rangle+(h_{z}+p_{0}E)\hat{\mathbf{z}}$ and $\boldsymbol{\mathcal{B}}_{B}=-Z\hat{\mathcal{J}}\cdot\langle\boldsymbol{\tau}_{A}\rangle+(h_{z}-p_{0}E)\hat{\mathbf{z}}$, where $
\hat{\mathcal J}=\operatorname{diag}(-J_\perp,-J_\perp,J_z)
$, such that the pseudospin exchange interaction is $\sum_{\langle ij\rangle}\btau_i\cdot\hat{\mathcal J}\cdot\btau_j$. As in a bipartite antiferromagnet~\cite{Andreev1980,Hals2011}, we introduce the collective variables, we introduce collective variables 
\begin{equation}
\cN=\frac{\langle\bm\tau_A\rangle-\langle\bm\tau_B\rangle}{2},\qquad \boldsymbol{\mathcal{M}}=\frac{\langle\bm\tau_A\rangle+\langle\bm\tau_B\rangle}{2} 
\end{equation}
where $\cN$ is the staggered pseudospin vector, hereafter referred to as the ``charge-order vector", while $\cM$ is its conjugate uniform component. Since $\vert \boldsymbol{\tau}_{A}\vert=\vert\boldsymbol{\tau}_{B} \vert=\tau_{0}$, $\boldsymbol{\mathcal{N}}\cdot\boldsymbol{\mathcal{M}}=0$ and $\cN^2+\cM^2=\tau_0^2$. The longitudinal component of $\cN$ and $\cM$ have direct physical interpretations: $\mathcal{N}_{z}$ acts as the charge-order parameter, while $\mathcal{M}_{z}=\frac{1}{2}(\bar{n}-1)$ measures the deviation from half filling, where $\bar{n}=(\langle n_{A}\rangle+\langle n_{B}\rangle)/2$ is the average electron occupation per site. At half filling ($h_{z}=0$), the two degenerate charge-ordered ground states are characterized by $\cM_{0}=0$ and $\cN_{0}=s\tau_{0}\hat{\mathbf{z}}$, where $s=\pm1$ labels the two opposite polarities of the charge order. Their transverse components, on the other hand, describe coherent mixing between the empty and doubly occupied local configurations and constitute the dynamical degrees of freedom associated with fluctuations of the longitudinal charge order.

\emph{Collective pseudospin excitations.---}Subtracting and adding the two sublattice equations in Eq. (\ref{eq:sublatticeEOM}) yields a pair of coupled antiferromagnetic-like equations of motion for $\cN$ and $\cM$, respectively. In the strong charge-ordering regime, where $|\cM|\ll|\cN|$, the conjugate component $\cM$ can be eliminated, yielding a closed equation of motion for the charge-order vector:
 \begin{align}
\cN\times&\Big\{I_{\mathcal{N}}\big[\ddot{\cN}+2\Omega_{z}\hat{\mathbf{z}}\times\dot{\cN}+\Omega_{z}\hat{\mathbf{z}}\times\big(\hat{\mathbf{z}}\times\cN\big)\big]\nonumber\\
&+\frac{\hbar\alpha_{G}}{\tau_{0}}\dot{\cN}
-\boldsymbol{\mathcal{B}}_{\mathcal{N}}\Big\}=0,
\label{eq:sigmamodel}
\end{align}   
where
\begin{equation}
   I_{\mathcal{N}}=\frac{\hbar^2}{Z(J_{z}-J_{\perp})\mathcal{N}_{0}^2+p_{0}{E}\mathcal{N}_{0}}
\end{equation}
is the effective inertia for the dynamics of the charge-order vector that is tunable by the applied electric field, with $\mathcal{N}_0=s\tau_0$, $\Omega_{z}=h_{z}/\hbar$ is the precession frequency induced by the uniform longitudinal field, and $
\boldsymbol{\mathcal{B}}_{\mathcal{N}}=-\partial\langle H_{\text{eff}}\rangle_{\text{MF}}/\partial\cN=Z\hat{\mathcal{J}}\cdot\cN+p_{0} E\hat{\mathbf{z}}$
is the effective field on the $\cN$ \cite{noteBN}, where $\langle H_{\text{eff}}\rangle_{\text{MF}}$ is the mean-field energy of the system. 

Eq.~(\ref{eq:sigmamodel}) constitutes a nonlinear sigma-model-type equation for the charge-order vector. In contrast to a purely relaxational scalar order parameter, $\cN$ exhibits intrinsic inertial dynamics, with its low-energy fluctuations predominantly transverse to the equilibrium $\cN_{0}$, i.e., $\vert \cN\vert\simeq \tau_{0}$, analogous to the dynamics of the N\'eel vector in antiferromagnets~\cite{Haldane1983,Hals2011,Tveten2013,Gomonay2014}. For small transverse fluctuations with $\boldsymbol{\eta}\cdot\boldsymbol{\mathcal{N}}_{0}=0$, substituting $\cN(t)=\cN_0+\boldsymbol\eta(t)$ into Eq.~(\ref{eq:sigmamodel}) leads to the linearized dynamic equation
\begin{equation}
\ddot{\eta}_{+}+(\gamma_{\text{CO}}+2i\Omega_{z})\dot{\eta}_{+}+(\Omega_{\text{CO}}^2-\Omega_{z}^2)\eta_{+}=0, 
\label{eq:dampedmode} 
\end{equation}
where $\eta_{+}=\eta_{x}+i\eta_{y}$ ($\eta_{-}=\eta_{+}^{\ast}$) represents the clockwise (anticlockwise) circularly polarized mode, $\gamma_{\text{CO}}=\hbar\alpha_{G}/(\tau_{0}I_{\mathcal{N}})$ is the decay rate, and
\begin{equation}
\Omega_\text{CO}^2=\frac{[p_{0}E+Z \mathcal{N}_{0}(J_z-J_\perp)][p_{0}E+Z\mathcal{N}_{0}(J_z+J_\perp)]}{\hbar^2}
\label{eq:omegaE}
\end{equation}
is the squared characteristic frequency of the transverse fluctuations in the absence of the uniform field $h_{z}$. Substituting $\boldsymbol{\eta}(t)\sim e^{-i\omega t}$ into Eq.~(\ref{eq:dampedmode}) yields the complex eigenfrequencies 
\begin{equation}
\omega_{\pm}=-\frac{i\gamma_{\text{CO}}}{2}+\sqrt{\Omega_{\text{CO}}^2-\frac{\gamma_{\text{CO}}^2}{4}\mp i \gamma_{\text{CO}}\Omega_{z}}\pm \Omega_{z}. \label{comw}
\end{equation}
In the absence of damping, $\omega_{\pm}$ reduces to $\Omega_{\text{CO}}\pm\Omega_{z}$, corresponding to the frequencies of the two circularly polarized transverse modes whose degeneracy is lifted by $h_z$. These collective modes closely resemble magnon excitations in antiferromagnets in the presence of an external magnetic field~\cite{rezende2020fundamentals,rezende2019introduction}. In contrast to antiferromagnetic magnons, however, the applied electric field can tune both $\Omega_{\text{CO}}$ and the decay rate $\gamma_{\text{CO}}$ through its modulation of the effective inertia, thereby providing direct electrical control over the collective charge-order dynamics. 

\begin{figure}
    \centering
    \includegraphics[width=8.6 cm]{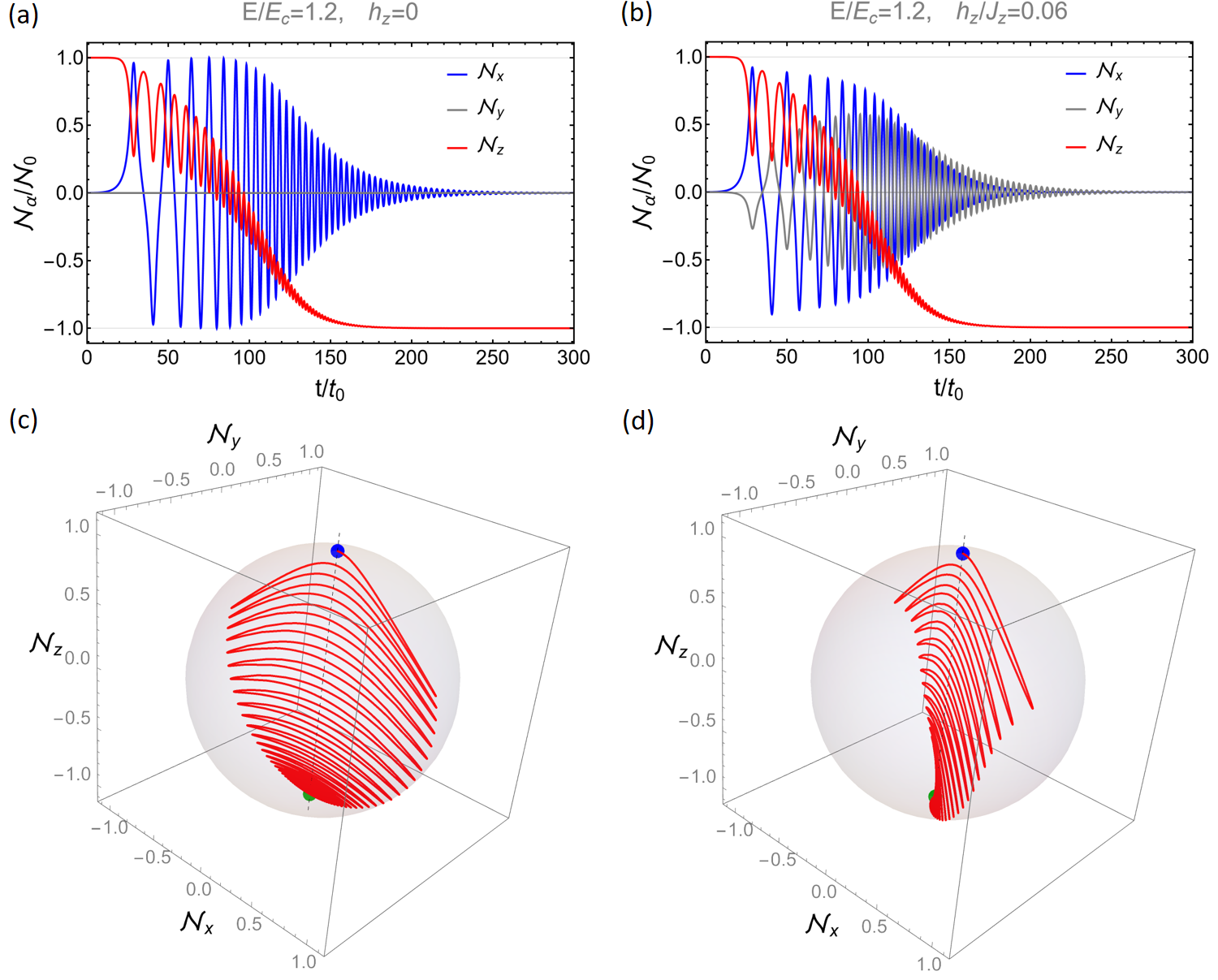}
    \caption{Time evolution and trajectories of the charge-order vector $\cN$ during
electric-field-driven switching of the charge-order polarity at $E/E_c=1.2$:
(a),(c) $h_z=0$ and (b),(d) $h_z/J_z=0.06$.
Here, time is measured in units of $t_0=\hbar/(ZJ_z\tau_0)$, with
$J_\perp/J_z=0.3$ and $\alpha_G=0.02$. The initial state is $\cN_0=\tau_0\hat{\mathbf{z}}$, corresponding to the positive-polarity CO state shown in Fig.~\ref{Fig-CO}.}
    \label{Fig-switching}
\end{figure}

\emph{Critical switching field.---}A sufficiently strong electric field can destabilize these collective modes and ultimately drive the reversal of the charge-order polarity, as illustrated in Fig.~\ref{Fig-CO}. The critical switching field $E_c$ is determined by the onset of instability of the transverse fluctuations $\boldsymbol{\eta}(t)\propto e^{-i\omega t}$ in Eq.~(\ref{eq:dampedmode}), which occurs when $\operatorname{Im}\omega_{\pm}=0$. In the absence of $h_{z}$, this condition yields
\begin{equation}
E_{c}=-\frac{Z\mathcal{N}_0(J_z-J_\perp)}{p_{0}}.
\label{eq:Ec}
\end{equation}
Since $J_z>J_\perp$, $E_c$ has the opposite sign to $\mathcal{N}0$ (assuming $p_0>0$), i.e., the switching field is antiparallel to the initial charge-order parameter. At $E=E_c$, $\gamma{\text{CO}}=0$ and $\Omega_{\text{CO}}^2=0$, indicating the softening of the transverse modes as their stiffness vanishes. For fields beyond the instability threshold, $\gamma_{\text{CO}}<0$ and $\Omega_{\text{CO}}^2<0$, such that transverse fluctuations grow exponentially and destabilize the initial charge-ordered state. As the charge-order polarity reverses from $\mathcal{N}_0$ to $-\mathcal{N}_0$, both $\gamma_{\text{CO}}>0$ and $\Omega_{\text{CO}}^2>0$ become positive again, indicating that the reversed charge-ordered state $-\mathcal{N}_0\hat{\mathbf z}$ is dynamically stable. Electric-field-induced switching can therefore be identified with the softening and dynamical instability of the transverse collective charge-pseudospin modes. Moreover, Eq.~(\ref{eq:Ec}) reveals reveals that the transverse $J_\perp$-terms in Eq.~(\ref{eq:Hps}) lowers the critical switching field by reducing the effective stiffness to $J_z-J_\perp$, whereas, in the absence of an electric field, the collective-mode frequency is governed by the geometric mean $\sqrt{(J_z-J_\perp)(J_z+J_\perp)}$. Measurements of both $\Omega_{\rm CO}$ and $E_c$ could therefore provide a means to separately determine $J_z$ and $J_\perp$. For a typical atomic-scale dipole moment $p_0\sim e\AA$, with $ZJ_z\simeq30~\mathrm{meV}$ and $ZJ_\perp\simeq10~\mathrm{meV}$, we estimate $|E_c|\simeq10^6~\mathrm{V/cm}$ and $\Omega_{\text{CO}}/2\pi\simeq3.4~\mathrm{THz}$, which are comparable to the intrinsic coercive fields of ferroelectrics~\cite{PhysRevLett.84.175,liu2016intrinsic} and the characteristic frequencies of the associated ferron excitations~\cite{Tang2022Ferron,Tang2024Ferron,Bauer2023Polarization,choe2026observation,zhang2026electricnat}, respectively. By contrast, the extremely low field of $\sim100~\mathrm{V/cm}$ recently observed for charge-order polarity switching in FeLuO$_3$~\cite{kikkawa2026electronic} is likely associated with the motion of charge-order domain walls~\cite{ma2015charge}, with the threshold field governed by extrinsic pinning barriers.

\emph{Numerical simulations.---}We numerically simulate the switching dynamics of the charge-order vector $\cN$ under an external electric field exceeding the critical value using Eq.~(\ref{eq:sublatticeEOM}), as shown in Fig.~\ref{Fig-switching}. The initial pseudospin states of two sublattices are taken as $\langle\boldsymbol{\tau}_{A}\rangle_{\mathrm{in}}=\tau_0(\epsilon,0,\sqrt{1-\epsilon^2})$ and $\langle\boldsymbol{\tau}_{B}\rangle_{\mathrm{in}}=-\langle\boldsymbol{\tau}_{A}\rangle_{\mathrm{in}}$, with a small transverse fluctuation $\epsilon=10^{-3}$ along the $x$ axis, corresponding to $\cN_{\mathrm{in}}=\tau_0(\epsilon,0,\sqrt{1-\epsilon^2})$ and $\cM_{\mathrm{in}}=0$; the switching dynamics, however, does not rely on the specific choice of the initial fluctuation configuration. For $h_z=0$, the dynamics of $\cN$ remains linearly polarized within the $xz$ plane. The longitudinal component $\mathcal{N}_z$, which represents the charge-order parameter, exhibits~\textit{oscillatory} dynamics during the switching from $+\tau_0\hat{\mathbf z}$ to $-\tau_0\hat{\mathbf z}$. Meanwhile, the transverse component $\mathcal{N}_x$ oscillates with an amplitude that initially grows rapidly due to the instability and subsequently decays as $\mathcal{N}_z$ approaches the reversed charge-ordered state $-\tau_0\hat{\mathbf z}$. In contrast, a finite $h_z$, which can arise from doping through a shift of the chemical potential, induces precession of $\cN$ around the $z$ axis in addition to the oscillatory reversal of $\mathcal{N}_z$. The resulting trajectory therefore exhibits nutational dynamics, similar to inertial magnetization dynamics in spin systems~\cite{PhysRevB.83.020410,neeraj2021inertial}.

\emph{Discussion.---}The field-driven charge-order dynamics developed here differs fundamentally from several familiar dynamical phenomena in charge-density-wave (CDW) and charge-ordered systems. The nonlinear electric-field response of CDW systems is predominantly associated with the depinning and translational sliding of the charge-density modulation~\cite{Gruner1988,Monceau2012}, for which the phase of the CDW constitutes the relevant collective coordinate. For an ideal incommensurate CDW in the absence of pinning, the corresponding phason is a gapless Goldstone mode associated with continuous translations of the density modulation~\cite{Gruner1988}. By contrast, the collective degree of freedom considered here is the local charge pseudospin formed by the $|0\rangle$ and $|2\rangle$ configurations. Its transverse excitation corresponds to a coherent rotation in pseudospin space rather than a translation of the charge-density modulation and is intrinsically gapped by the pseudospin anisotropy $J_z-J_\perp$. 

The present pseudospin dynamics should also be distinguished from the extensively investigated photoinduced melting of charge order~\cite{fiebig2000sub,PhysRevB.63.113105,Iwai2007,Yonemitsu2007,Matsuzaki2009,PhysRevLett.103.155702,Yada2016}, in which optical excitation transiently suppresses the charge-ordered state, often accompanied by substantial electronic and structural reorganization. Here, instead,
an electric field reverses the charge-order polarity by exerting a pseudospin
torque that drives a coherent rotation of the staggered charge-order vector
$\cN$ in pseudospin space, reminiscent of current-induced order-parameter
switching in magnetic systems~\cite{Katine2000,Miron2011Nature,Liu2012Science}. Our findings identify a regime of coherent electrically driven charge-order dynamics described in terms of pseudospins as collective dynamical degrees of freedom, distinct from conventional CDW sliding and photoinduced charge-order melting.

\emph{Conclusions.---}We derived an effective pseudospin Hamiltonian for the collective dynamics of charge order from an extended Hubbard model, in which the charge-ordered state corresponds to staggered pseudospin order. The resulting pseudospin dynamics of charge order closely resembles that of the N\'eel vector in a bipartite antiferromagnet. We show that an external electric field can directly tune the frequency and damping of the collective pseudospin modes and, above a critical field, drive a dynamical instability that enables coherent reversal of the charge-order polarity. This provides a charge-sector analogue of spin manipulation in magnets, with electric fields serving as a natural control parameter. Time-resolved measurements of charge disproportionation under strong electric-field pulses could directly probe the predicted coherent switching dynamics. The close correspondence between charge-pseudospin and antiferromagnetic dynamics suggests that concepts developed for the dynamical control of magnetic order may have direct counterparts in the charge sector, opening a route toward coherent electrical manipulation of charge order and extending the paradigm of spin-order dynamics in spintronics to the charge sector.

\emph{Acknowledgments.---}
This work was supported by JSPS KAKENHI Grant-in-Aid for Scientific Research (B) (Grant No. 26K00625).

\bibliography{reference}

@misc{SM,
    note ={See the Supplemental Material for a detailed derivation of the effective charge-pseudospin Hamiltonian in Eq.~(\ref{eq:Hps}) from the extended Hubbard model.}
}

@misc{notedipole,
    key="{In a conducting charge-ordered state}",
    note ="{the intracell electric dipole is screened by conduction electrons, preventing an external electric field from coupling directly to the charge order. Instead, an effective coupling between the electric field and the charge order is mediated by the nontrivial geometric properties of the conduction electrons~\cite{kikkawa2026electronic}.}"
}

@misc{noteBN,
    key="{In the macropseudospin approximation}",
    note ="{the charge-order dynamics is assumed to be spatially uniform, such that gradient terms in the staggered effective field $\boldsymbol{\mathcal{B}}_{\mathcal{N}}$ are absent. For spatially nonuniform dynamics, an exchange-stiffness contribution proportional to $\nabla^2\cN$ arises in $\boldsymbol{\mathcal{B}}_{\mathcal{N}}$.}"
}

@misc{energy,
    key="{The energy cost $\Delta$ depends on the local environment of the two nearest-neighbor sites. For a uniform charge-ordered ground state}", 
    note = "{the energy cost for creating the intermediate virtual state with two singly occupied sites is $\Delta=(2Z-1)V-U$, assuming all other sites are fixed in their ground-state configurations.}"
}

@article{Schrieffer1966,
  author = {Schrieffer, J. R. and Wolff, P. A.},
  title = {Relation between the Anderson and Kondo Hamiltonians},
  journal = {Phys. Rev.},
  volume = {149},
  pages = {491--492},
  year = {1966},
  doi = {10.1103/PhysRev.149.491}
}

@article{Takahashi1977,
  author = {Takahashi, M.},
  title = {Half-filled Hubbard model at low temperature},
  journal = {J. Phys. C: Solid State Phys.},
  volume = {10},
  pages = {1289--1301},
  doi={10.1088/0022-3719/10/8/031},
  year = {1977}
}

@article{Gruner1988,
  author  = {Gr{\"u}ner, George},
  title   = {The Dynamics of Charge-Density Waves},
  journal = {Rev. Mod. Phys.},
  volume  = {60},
  pages   = {1129--1181},
  year    = {1988},
  doi     = {10.1103/RevModPhys.60.1129}
}

@article{Monceau2012,
  author  = {Monceau, Pierre},
  title   = {Electronic Crystals: An Experimental Overview},
  journal = {Adv. Phys.},
  volume  = {61},
  pages   = {325--581},
  year    = {2012},
  doi     = {10.1080/00018732.2012.719674}
}

@article{Iwai2007,
  author  = {Iwai, S. and Yamamoto, K. and Kashiwazaki, A. and
             Hiramatsu, F. and Nakaya, H. and Kawakami, Y. and
             Yakushi, K. and Okamoto, H. and Mori, H. and Nishio, Y.},
  title   = {Photoinduced Melting of a Stripe-Type Charge-Order and
             Metallic Domain Formation in a Layered {BEDT-TTF}-Based
             Organic Salt},
  journal = {Phys. Rev. Lett.},
  volume  = {98},
  pages   = {097402},
  year    = {2007},
  doi     = {10.1103/PhysRevLett.98.097402}}

@article{Matsuzaki2009,
  author  = {Matsuzaki, Hiroyuki and Uemura, Hirotaka and Matsubara, Masakazu
             and Kimura, Tsuyoshi and Tokura, Yoshinori and Okamoto, Hiroshi},
  title   = {Detecting Charge and Lattice Dynamics in Photoinduced
             Charge-Order Melting in Perovskite-Type Manganites
             Using a 30-Femtosecond Time Resolution},
  journal = {Phys. Rev. B},
  volume  = {79},
  pages   = {235131},
  year    = {2009},
  doi     = {10.1103/PhysRevB.79.235131}}

@article{Yonemitsu2007,
  author  = {Yonemitsu, Kenji and Maeshima, Nobuya},
  title   = {Photoinduced Melting of Charge Order in a Quarter-Filled
             Electron System Coupled with Different Types of Phonons},
  journal = {Phys. Rev. B},
  volume  = {76},
  pages   = {075105},
  year    = {2007},
  doi     = {10.1103/PhysRevB.76.075105}
}

@article{PhysRevB.63.113105,
  title = {Dynamics of photoinduced melting of charge/orbital order in a layered manganite ${\mathrm{La}}_{0.5}{\mathrm{Sr}}_{1.5}{\mathrm{MnO}}_{4}$},
  author = {Ogasawara, T. and Kimura, T. and Ishikawa, T. and Kuwata-Gonokami, M. and Tokura, Y.},
  journal = {Phys. Rev. B},
  volume = {63},
  issue = {11},
  pages = {113105},
  numpages = {4},
  year = {2001},
  month = {Mar},
  publisher = {American Physical Society},
  doi = {10.1103/PhysRevB.63.113105},
  url = {https://link.aps.org/doi/10.1103/PhysRevB.63.113105}
}

@article{Yada2016,
  author  = {Yada, H. and Ijiri, Y. and Uemura, H. and
             Tomioka, Y. and Okamoto, H.},
  title   = {Enhancement of Photoinduced Charge-Order Melting via
             Anisotropy Control by Double-Pulse Excitation in
             Perovskite Manganites: {Pr}$_{0.6}${Ca}$_{0.4}${MnO}$_3$},
  journal = {Phys. Rev. Lett.},
  volume  = {116},
  pages   = {076402},
  year    = {2016},
  doi     = {10.1103/PhysRevLett.116.076402}
}

@article{MacDonald1988,
  author = {MacDonald, A. H. and Girvin, S. M. and Yoshioka, D.},
  title = {$t/U$ expansion for the Hubbard model},
  journal = {Phys. Rev. B},
  volume = {37},
  pages = {9753--9756},
  year = {1988},
  doi = {10.1103/PhysRevB.37.9753}
}

@article{Bravyi2011,
  author = {Bravyi, Sergey and DiVincenzo, David P. and Loss, Daniel},
  title = {Schrieffer-Wolff transformation for quantum many-body systems},
  journal = {Ann. Phys.},
  volume = {326},
  pages = {2793--2826},
  year = {2011},
  doi = {10.1016/j.aop.2011.06.004}
}

@article{Andreev1980,
  author = {Andreev, A. F. and Marchenko, V. I.},
  title = {Symmetry and the Macroscopic Dynamics of Magnetic Materials},
  journal = {Sov. Phys. Usp.},
  volume = {23},
  pages = {21--34},
  year = {1980},
  doi = {10.1070/PU1980v023n01ABEH004859}
}

@article{Tveten2013,
  author = {Tveten, E. G. and Qaiumzadeh, A. and Tretiakov, O. A. and Brataas, A.},
  title = {Staggered Dynamics in Antiferromagnets by Collective Coordinates},
  journal = {Phys. Rev. Lett.},
  volume = {110},
  pages = {127208},
  year = {2013},
  doi = {10.1103/PhysRevLett.110.127208}
}

@article{Gomonay2014,
  author = {Gomonay, H. V. and Loktev, V. M.},
  title = {Spintronics of Antiferromagnetic Systems},
  journal = {Low Temp. Phys.},
  volume = {40},
  pages = {17--35},
  year = {2014},
  doi = {10.1063/1.4862467}
}

@article{Hals2011,
  author = {Hals, Kjetil M. D. and Tserkovnyak, Yaroslav and Brataas, Arne},
  title = {Phenomenology of Current-Induced Dynamics in Antiferromagnets},
  journal = {Phys. Rev. Lett.},
  volume = {106},
  pages = {107206},
  year = {2011},
  doi = {10.1103/PhysRevLett.106.107206}
}

@book{rezende2020fundamentals,
  title={Fundamentals of magnonics},
  author={Rezende, Sergio M},
  volume={969},
  year={2020},
  publisher={Springer}
}

@article{PhysRevLett.84.175,
  title = {Intrinsic Ferroelectric Coercive Field},
  author = {Ducharme, Stephen and Fridkin, V. M. and Bune, A. V. and Palto, S. P. and Blinov, L. M. and Petukhova, N. N. and Yudin, S. G.},
  journal = {Phys. Rev. Lett.},
  volume = {84},
  issue = {1},
  pages = {175--178},
  numpages = {0},
  year = {2000},
  month = {Jan},
  publisher = {American Physical Society},
  doi = {10.1103/PhysRevLett.84.175},
  url = {https://link.aps.org/doi/10.1103/PhysRevLett.84.175}
}

@article{Tang2022Ferron,
  title = {Excitations of the Ferroelectric Order},
  author = {Tang, Ping and Iguchi, Ryo and Uchida, Ken-ichi and Bauer, Gerrit E. W.},
  journal = {Phys. Rev. B},
  volume = {106},
  pages = {L081105},
  year = {2022},
  doi = {10.1103/PhysRevB.106.L081105}
}

@article{ma2015charge,
  title={Charge-order domain walls with enhanced conductivity in a layered manganite},
  author={Ma, Eric Yue and Bryant, Benjamin and Tokunaga, Yusuke and Aeppli, Gabriel and Tokura, Yoshinori and Shen, Zhi-Xun},
  journal={Nature Communications},
  volume={6},
  number={1},
  pages={7595},
  year={2015},
  url={https://doi.org/10.1038/ncomms8595},
  publisher={Nature Publishing Group UK London}
}

@article{PhysRevB.83.020410,
  title = {Magnetization dynamics in the inertial regime: Nutation predicted at short time scales},
  author = {Ciornei, M.-C. and Rub\'{\i}, J. M. and Wegrowe, J.-E.},
  journal = {Phys. Rev. B},
  volume = {83},
  issue = {2},
  pages = {020410(R)},
  numpages = {4},
  year = {2011},
  month = {Jan},
  publisher = {American Physical Society},
  doi = {10.1103/PhysRevB.83.020410},
  url = {https://link.aps.org/doi/10.1103/PhysRevB.83.020410}
}

@article{neeraj2021inertial,
  title={Inertial spin dynamics in ferromagnets},
  author={Neeraj, Kumar and Awari, Nilesh and Kovalev, Sergey and Polley, Debanjan and Zhou Hagstr{\"o}m, Nanna and Arekapudi, Sri Sai Phani Kanth and Semisalova, Anna and Lenz, Kilian and Green, Bertram and Deinert, Jan-Christoph and others},
  journal={Nature Physics},
  volume={17},
  number={2},
  pages={245--250},
  year={2021},
  url={https://doi.org/10.1038/s41567-020-01040-y},
  publisher={Nature Publishing Group UK London}
}

@article{Bauer2023Polarization,
  title = {Polarization Transport in Ferroelectrics},
  author = {Bauer, Gerrit E. W. and Tang, Ping and Iguchi, Ryo and Xiao, Jiang
            and Shen, Ka and Zhong, Zhiyong and Yu, Tao and Rezende, Sergio M.
            and Heremans, Joseph P. and Uchida, Ken-ichi},
  journal = {Phys. Rev. Applied},
  volume = {20},
  pages = {050501},
  year = {2023},
  doi = {10.1103/PhysRevApplied.20.050501}
}

@article{liu2016intrinsic,
  title={Intrinsic ferroelectric switching from first principles},
  author={Liu, Shi and Grinberg, Ilya and Rappe, Andrew M},
  journal={Nature},
  volume={534},
  number={7607},
  pages={360--363},
  year={2016},
  url={https://doi.org/10.1038/nature18286},
  publisher={Nature Publishing Group UK London}
}

@article{Tang2024Ferron,
  title = {Electric Analog of Magnons in Order-Disorder Ferroelectrics},
  author = {Tang, Ping and Bauer, Gerrit E. W.},
  journal = {Phys. Rev. B},
  volume = {109},
  pages = {L060301},
  year = {2024},
  doi = {10.1103/PhysRevB.109.L060301}
}

@article{rezende2019introduction,
  title={Introduction to antiferromagnetic magnons},
  author={Rezende, Sergio M and Azevedo, Antonio and Rodr{\'\i}guez-Su{\'a}rez, Roberto L},
  journal={Journal of Applied Physics},
  volume={126},
  number={15},
  year={2019},
  url={https://doi.org/10.1063/1.5109132},
  publisher={AIP Publishing}
}

@article{Haldane1983,
  author = {Haldane, F. D. M.},
  title = {Nonlinear Field Theory of Large-Spin Heisenberg
           Antiferromagnets: Semiclassically Quantized Solitons
           of the One-Dimensional Easy-Axis N\'eel State},
  journal = {Phys. Rev. Lett.},
  volume = {50},
  pages = {1153--1156},
  year = {1983},
  doi = {10.1103/PhysRevLett.50.1153}
}

@article{Attfield2006,
  author  = {Attfield, J. Paul},
  title   = {Charge Ordering in Transition Metal Oxides},
  journal = {Solid State Sci.},
  volume  = {8},
  pages   = {861--867},
  year    = {2006},
  doi     = {10.1016/j.solidstatesciences.2005.02.011}
}

@article{PhysRevLett.105.187401,
  title = {Ultrafast Melting of a Charge-Density Wave in the Mott Insulator $1T\mathrm{\text{\ensuremath{-}}}{\mathrm{TaS}}_{2}$},
  author = {Hellmann, S. and Beye, M. and Sohrt, C. and Rohwer, T. and Sorgenfrei, F. and Redlin, H. and Kall\"ane, M. and Marczynski-B\"uhlow, M. and Hennies, F. and Bauer, M. and F\"ohlisch, A. and Kipp, L. and Wurth, W. and Rossnagel, K.},
  journal = {Phys. Rev. Lett.},
  volume = {105},
  issue = {18},
  pages = {187401},
  numpages = {4},
  year = {2010},
  month = {Oct},
  publisher = {American Physical Society},
  doi = {10.1103/PhysRevLett.105.187401},
  url = {https://link.aps.org/doi/10.1103/PhysRevLett.105.187401}
}

@article{fiebig2000sub,
  title={Sub-picosecond photo-induced melting of a charge-ordered state in a perovskite manganite},
  author={Fiebig, Manfred and Miyano, Kenjiro and Tomioka, Yoshinori and Tokura, Yoshinori},
  journal={Applied Physics B},
  volume={71},
  number={2},
  url={https://doi.org/10.1007/s003400000338},
  pages={211--215},
  year={2000},
  publisher={Springer}
}

@article{PhysRevLett.103.155702,
  title = {Ultrafast Structural Phase Transition Driven by Photoinduced Melting of Charge and Orbital Order},
  author = {Beaud, P. and Johnson, S. L. and Vorobeva, E. and Staub, U. and Souza, R. A. De and Milne, C. J. and Jia, Q. X. and Ingold, G.},
  journal = {Phys. Rev. Lett.},
  volume = {103},
  issue = {15},
  pages = {155702},
  numpages = {4},
  year = {2009},
  month = {Oct},
  publisher = {American Physical Society},
  doi = {10.1103/PhysRevLett.103.155702},
  url = {https://link.aps.org/doi/10.1103/PhysRevLett.103.155702}
}

@article{PhysRevB.70.235107,
  title = {Charge ordering in extended Hubbard models: Variational cluster approach},
  author = {Aichhorn, M. and Evertz, H. G. and von der Linden, W. and Potthoff, M.},
  journal = {Phys. Rev. B},
  volume = {70},
  issue = {23},
  pages = {235107},
  numpages = {13},
  year = {2004},
  month = {Dec},
  publisher = {American Physical Society},
  doi = {10.1103/PhysRevB.70.235107},
  url = {https://link.aps.org/doi/10.1103/PhysRevB.70.235107}
}

@article{ishihara2010electronic,
  title={Electronic ferroelectricity and frustration},
  author={Ishihara, Sumio},
  journal={Journal of the Physical Society of Japan},
  volume={79},
  number={1},
  pages={011010},
  year={2010},
  url={https://journals.jps.jp/doi/citedby/10.1143/JPSJ.79.011010},
  publisher={The Physical Society of Japan}
}

@article{PhysRevB.54.17452,
  title = {Theory of electronic ferroelectricity},
  author = {Portengen, T. and \"Ostreich, Th. and Sham, L. J.},
  journal = {Phys. Rev. B},
  volume = {54},
  issue = {24},
  pages = {17452--17463},
  numpages = {0},
  year = {1996},
  month = {Dec},
  publisher = {American Physical Society},
  doi = {10.1103/PhysRevB.54.17452},
  url = {https://link.aps.org/doi/10.1103/PhysRevB.54.17452}
}

@article{Seo2000,
  title = {Charge Ordering in Organic {ET} Compounds},
  author = {Seo, Hitoshi},
  journal = {J. Phys. Soc. Jpn.},
  volume = {69},
  number = {3},
  pages = {805--820},
  year = {2000},
  doi = {10.1143/JPSJ.69.805}
}

@article{PhysRevB.49.9670,
  title = {Phase diagram of the one-dimensional extended Hubbard model with attractive and/or repulsive interactions at quarter filling},
  author = {Penc, Karlo and Mila, Fr\'ed\'eric},
  journal = {Phys. Rev. B},
  volume = {49},
  issue = {14},
  pages = {9670--9678},
  numpages = {0},
  year = {1994},
  month = {Apr},
  publisher = {American Physical Society},
  doi = {10.1103/PhysRevB.49.9670},
  url = {https://link.aps.org/doi/10.1103/PhysRevB.49.9670}
}

@article{PhysRevB.19.3970,
  title = {Electric field depinning of charge density waves},
  author = {Lee, P. A. and Rice, T. M.},
  journal = {Phys. Rev. B},
  volume = {19},
  issue = {8},
  pages = {3970--3980},
  numpages = {0},
  year = {1979},
  month = {Apr},
  publisher = {American Physical Society},
  doi = {10.1103/PhysRevB.19.3970},
  url = {https://link.aps.org/doi/10.1103/PhysRevB.19.3970}
}

@article{Seo2004,
  author  = {Seo, Hitoshi and Hotta, Chisa and Fukuyama, Hidetoshi},
  title   = {Toward Systematic Understanding of Diversity of Electronic Properties in Low-Dimensional Molecular Solids},
  journal = {Chem. Rev.},
  volume  = {104},
  pages   = {5005--5036},
  year    = {2004},
  doi     = {10.1021/cr030646k}
}

@article{Senn2012,
  author  = {Senn, Mark S. and Wright, Jon P. and Attfield, J. Paul},
  title   = {Charge Order and Three-Site Distortions in the {Verwey} Structure of Magnetite},
  journal = {Nature},
  volume  = {481},
  pages   = {173--176},
  year    = {2012},
  doi     = {10.1038/nature10704}
}

@article{Paki2019,
  author  = {Paki, Joseph and Terletska, Hanna and Iskakov, Sergei and Gull, Emanuel},
  title   = {Charge Order and Antiferromagnetism in the Extended {Hubbard} Model},
  journal = {Phys. Rev. B},
  volume  = {99},
  pages   = {245146},
  year    = {2019},
  doi     = {10.1103/PhysRevB.99.245146}
}

@article{Stojchevska2014,
  author  = {Stojchevska, L. and Vaskivskyi, I. and Mertelj, T. and Ku{\v{s}}ar, P. and Svetin, D. and Brazovskii, S. and Mihailovic, D.},
  title   = {Ultrafast Switching to a Stable Hidden Quantum State in an Electronic Crystal},
  journal = {Science},
  volume  = {344},
  pages   = {177--180},
  year    = {2014},
  doi     = {10.1126/science.1241591}
}

@article{Geremew2019,
  author  = {Geremew, Adane K. and Rumyantsev, Sergey and Kargar, Fariborz and Debnath, Bishwajit and Nosek, Adrian and Bloodgood, Matthew A. and Bockrath, Marc and Salguero, Tina T. and Lake, Roger K. and Balandin, Alexander A.},
  title   = {Bias-Voltage Driven Switching of the Charge-Density-Wave and Normal Metallic Phases in 1T-{TaS}$_2$ Thin-Film Devices},
  journal = {ACS Nano},
  volume  = {13},
  pages   = {7231--7240},
  year    = {2019},
  doi     = {10.1021/acsnano.9b02870}
}

@article{Fert2008,
  author  = {Fert, Albert},
  title   = {Nobel Lecture: Origin, Development, and Future of Spintronics},
  journal = {Rev. Mod. Phys.},
  volume  = {80},
  pages   = {1517--1530},
  year    = {2008},
  doi     = {10.1103/RevModPhys.80.1517}
}

@article{Landau1935,
  author  = {Landau, L. D. and Lifshitz, E. M.},
  title   = {On the Theory of the Dispersion of Magnetic Permeability
             in Ferromagnetic Bodies},
  journal = {Phys. Z. Sowjetunion},
  volume  = {8},
  pages   = {153--169},
  year    = {1935}
}

@article{Gilbert2004,
  author  = {Gilbert, T. L.},
  title   = {A Phenomenological Theory of Damping in Ferromagnetic Materials},
  journal = {IEEE Trans. Magn.},
  volume  = {40},
  pages   = {3443--3449},
  year    = {2004},
  doi     = {10.1109/TMAG.2004.836740}
}

@article{Slonczewski1996,
  author  = {Slonczewski, J. C.},
  title   = {Current-Driven Excitation of Magnetic Multilayers},
  journal = {J. Magn. Magn. Mater.},
  volume  = {159},
  pages   = {L1--L7},
  year    = {1996},
  doi     = {10.1016/0304-8853(96)00062-5}
}

@article{Berger1996,
  author  = {Berger, L.},
  title   = {Emission of Spin Waves by a Magnetic Multilayer Traversed
             by a Current},
  journal = {Phys. Rev. B},
  volume  = {54},
  pages   = {9353--9358},
  year    = {1996},
  doi     = {10.1103/PhysRevB.54.9353}
}

@article{Ralph2008,
  author  = {Ralph, D. C. and Stiles, M. D.},
  title   = {Spin Transfer Torques},
  journal = {J. Magn. Magn. Mater.},
  volume  = {320},
  pages   = {1190--1216},
  year    = {2008},
  doi     = {10.1016/j.jmmm.2007.12.019}
}

@article{kikkawa2026electronic,
  title={Electronic manipulation of polar order in electron crystal},
  author={Kikkawa, Takashi and Chen, Ziyan and Fujimoto, Yuto and Morimoto, Takahiro and Hirata, Yuya and Arisawa, Hiroki and Kaverzin, Alexey A and Okamoto, Satoshi and Okimoto, Yoichi and Ikeda, Naoshi and others},
  journal={arXiv preprint arXiv:2607.09425},
  url={https://doi.org/10.48550/arXiv.2607.09425},
  year={2026}
}

@article{comin2016resonant,
  title={Resonant x-ray scattering studies of charge order in cuprates},
  author={Comin, Riccardo and Damascelli, Andrea},
  journal={Annual Review of Condensed Matter Physics},
  volume={7},
  number={1},
  pages={369--405},
  year={2016},
  url={https://www.annualreviews.org/content/journals/10.1146/annurev-conmatphys-031115-011401},
  publisher={Annual Reviews}
}

@article{takahashi2006charge,
  title={Charge ordering in organic conductors},
  author={Takahashi, Toshihiro and Nogami, Yoshio and Yakushi, Kyuya},
  journal={journal of the physical society of japan},
  volume={75},
  number={5},
  pages={051008--051008},
  year={2006},
  url={https://doi.org/10.1143/jpsj.75.051008},
  publisher={The Physical Society of Japan}
}

@article{renner2002atomic,
  title={Atomic-scale images of charge ordering in a mixed-valence manganite},
  author={Renner, Ch and Aeppli, G and Kim, B-G and Soh, Yeong-Ah and Cheong, S-W},
  journal={Nature},
  volume={416},
  number={6880},
  pages={518--521},
  year={2002},
  url={https://doi.org/10.1038/416518a},
  publisher={Nature Publishing Group UK London}
}

@article{PhysRevB.81.134417,
  title = {Charge and spin ordering in the mixed-valence compound ${\text{LuFe}}_{2}{\text{O}}_{4}$},
  author = {Harris, A. B. and Yildirim, T.},
  journal = {Phys. Rev. B},
  volume = {81},
  issue = {13},
  pages = {134417},
  numpages = {15},
  year = {2010},
  month = {Apr},
  publisher = {American Physical Society},
  doi = {10.1103/PhysRevB.81.134417},
  url = {https://link.aps.org/doi/10.1103/PhysRevB.81.134417}
}

@article{PhysRevB.62.12167,
  title = {Charge and spin ordering process in the mixed-valence system ${\mathrm{LuFe}}_{2}{\mathrm{O}}_{4}:$ Charge ordering},
  author = {Yamada, Y. and Kitsuda, K. and Nohdo, S. and Ikeda, N.},
  journal = {Phys. Rev. B},
  volume = {62},
  issue = {18},
  pages = {12167--12174},
  numpages = {0},
  year = {2000},
  month = {Nov},
  publisher = {American Physical Society},
  doi = {10.1103/PhysRevB.62.12167},
  url = {https://link.aps.org/doi/10.1103/PhysRevB.62.12167}
}

@article{flicker2015charge,
  title={Charge order from orbital-dependent coupling evidenced by NbSe2},
  author={Flicker, Felix and Van Wezel, Jasper},
  journal={Nature communications},
  volume={6},
  number={1},
  pages={7034},
  year={2015},
  url={https://doi.org/10.1038/ncomms8034},
  publisher={Nature Publishing Group UK London}
}

@article{hwang2024charge,
  title={Charge density waves in two-dimensional transition metal dichalcogenides},
  author={Hwang, Jinwoong and Ruan, Wei and Chen, Yi and Tang, Shujie and Crommie, Michael F and Shen, Zhi-Xun and Mo, Sung-Kwan},
  journal={Reports on Progress in Physics},
  volume={87},
  number={4},
  pages={044502},
  year={2024},
  doi={10.1088/1361-6633/ad36d3},
  publisher={IOP Publishing}
}

@article{PhysRevLett.82.4679,
  title = {Charge Ordering and Long-Range Interactions in Layered Transition Metal Oxides},
  author = {Stojkovi\ifmmode \acute{c}\else \'{c}\fi{}, Branko P. and Yu, Z. G. and Bishop, A. R. and Neto, A. H. Castro and Gr\o{}nbech-Jensen, Niels},
  journal = {Phys. Rev. Lett.},
  volume = {82},
  issue = {23},
  pages = {4679--4682},
  numpages = {0},
  year = {1999},
  month = {Jun},
  publisher = {American Physical Society},
  doi = {10.1103/PhysRevLett.82.4679},
  url = {https://link.aps.org/doi/10.1103/PhysRevLett.82.4679}
}

@article{tokura2000orbital,
  title={Orbital physics in transition-metal oxides},
  author={Tokura, Yoshinori and Nagaosa, Naoto},
  journal={Science},
  volume={288},
  number={5465},
  pages={462--468},
  year={2000},
  doi={10.1126/science.288.5465.462},
  publisher={American Association for the Advancement of Science}
}

@article{PhysRevB.62.R7679,
  title = {Charge ordering in a quasi-two-dimensional organic conductor},
  author = {Miyagawa, K. and Kawamoto, A. and Kanoda, K.},
  journal = {Phys. Rev. B},
  volume = {62},
  issue = {12},
  pages = {R7679(R)--R7682(R)},
  numpages = {0},
  year = {2000},
  month = {Sep},
  publisher = {American Physical Society},
  doi = {10.1103/PhysRevB.62.R7679},
  url = {https://link.aps.org/doi/10.1103/PhysRevB.62.R7679}
}

@article{dressel2004optical,
  title={Optical properties of two-dimensional organic conductors: Signatures of charge ordering and correlation effects},
  author={Dressel, Martin and Drichko, Natalia},
  journal={Chemical Reviews},
  volume={104},
  number={11},
  pages={5689--5716},
  year={2004},
  url={https://doi.org/10.1021/cr030642f},
  publisher={ACS Publications}
}

@article{RevModPhys.70.1039,
  title = {Metal-insulator transitions},
  author = {Imada, Masatoshi and Fujimori, Atsushi and Tokura, Yoshinori},
  journal = {Rev. Mod. Phys.},
  volume = {70},
  issue = {4},
  pages = {1039--1263},
  numpages = {0},
  year = {1998},
  month = {Oct},
  publisher = {American Physical Society},
  doi = {10.1103/RevModPhys.70.1039},
  url = {https://link.aps.org/doi/10.1103/RevModPhys.70.1039}
}

@article{RevModPhys.87.457,
  title = {Colloquium: Theory of intertwined orders in high temperature superconductors},
  author = {Fradkin, Eduardo and Kivelson, Steven A. and Tranquada, John M.},
  journal = {Rev. Mod. Phys.},
  volume = {87},
  issue = {2},
  pages = {457--482},
  numpages = {26},
  year = {2015},
  month = {May},
  publisher = {American Physical Society},
  doi = {10.1103/RevModPhys.87.457},
  url = {https://link.aps.org/doi/10.1103/RevModPhys.87.457}
}

@article{comin2015symmetry,
  title={Symmetry of charge order in cuprates},
  author={Comin, R and Sutarto, R and He, F and da Silva Neto, EH and Chauviere, L and Frano, A and Liang, R and Hardy, WN and Bonn, DA and Yoshida, Y and others},
  journal={Nature Materials},
  volume={14},
  number={8},
  pages={796--800},
  year={2015},
  url={https://doi.org/10.1038/nmat4295},
  publisher={Nature Publishing Group UK London}
}

@article{PhysRevB.95.115149,
  title = {Charge ordering and correlation effects in the extended Hubbard model},
  author = {Terletska, Hanna and Chen, Tianran and Gull, Emanuel},
  journal = {Phys. Rev. B},
  volume = {95},
  issue = {11},
  pages = {115149},
  numpages = {11},
  year = {2017},
  month = {Mar},
  publisher = {American Physical Society},
  doi = {10.1103/PhysRevB.95.115149},
  url = {https://link.aps.org/doi/10.1103/PhysRevB.95.115149}
}

@article{PhysRevLett.111.036601,
  title = {Optimal Hubbard Models for Materials with Nonlocal Coulomb Interactions: Graphene, Silicene, and Benzene},
  author = {Sch\"uler, M. and R\"osner, M. and Wehling, T. O. and Lichtenstein, A. I. and Katsnelson, M. I.},
  journal = {Phys. Rev. Lett.},
  volume = {111},
  issue = {3},
  pages = {036601},
  numpages = {5},
  year = {2013},
  month = {Jul},
  publisher = {American Physical Society},
  doi = {10.1103/PhysRevLett.111.036601},
  url = {https://link.aps.org/doi/10.1103/PhysRevLett.111.036601}
}

@article{PhysRevLett.110.166401,
  title = {Long-Range Coulomb Interactions in Surface Systems: A First-Principles Description within Self-Consistently Combined $GW$ and Dynamical Mean-Field Theory},
  author = {Hansmann, P. and Ayral, T. and Vaugier, L. and Werner, P. and Biermann, S.},
  journal = {Phys. Rev. Lett.},
  volume = {110},
  issue = {16},
  pages = {166401},
  numpages = {5},
  year = {2013},
  month = {Apr},
  publisher = {American Physical Society},
  doi = {10.1103/PhysRevLett.110.166401},
  url = {https://link.aps.org/doi/10.1103/PhysRevLett.110.166401}
}

@article{PhysRevLett.106.236805,
  title = {Strength of Effective Coulomb Interactions in Graphene and Graphite},
  author = {Wehling, T. O. and \ifmmode \mbox{\c{S}}\else \c{S}\fi{}a\ifmmode \mbox{\c{s}}\else \c{s}\fi{}\ifmmode \imath \else \i \fi{}o\ifmmode \breve{g}\else \u{g}\fi{}lu, E. and Friedrich, C. and Lichtenstein, A. I. and Katsnelson, M. I. and Bl\"ugel, S.},
  journal = {Phys. Rev. Lett.},
  volume = {106},
  issue = {23},
  pages = {236805},
  numpages = {4},
  year = {2011},
  month = {Jun},
  publisher = {American Physical Society},
  doi = {10.1103/PhysRevLett.106.236805},
  url = {https://link.aps.org/doi/10.1103/PhysRevLett.106.236805}
}

@article{PhysRevLett.53.2327,
  title = {Charge-Density-Wave to Spin-Density-Wave Transition in the Extended Hubbard Model},
  author = {Hirsch, J. E.},
  journal = {Phys. Rev. Lett.},
  volume = {53},
  issue = {24},
  pages = {2327--2330},
  numpages = {0},
  year = {1984},
  month = {Dec},
  publisher = {American Physical Society},
  doi = {10.1103/PhysRevLett.53.2327},
  url = {https://link.aps.org/doi/10.1103/PhysRevLett.53.2327}
}

@article{PhysRevB.39.9397,
  title = {Extended Hubbard model in two dimensions},
  author = {Zhang, Y. and Callaway, J.},
  journal = {Phys. Rev. B},
  volume = {39},
  issue = {13},
  pages = {9397--9404},
  numpages = {0},
  year = {1989},
  month = {May},
  publisher = {American Physical Society},
  doi = {10.1103/PhysRevB.39.9397},
  url = {https://link.aps.org/doi/10.1103/PhysRevB.39.9397}
}

@article{Oles2010,
  author  = {Ole{\'s}, Andrzej M.},
  title   = {Charge and Orbital Order in Transition Metal Oxides},
  journal = {Acta Physica Polonica A},
  volume  = {118},
  number  = {2},
  pages   = {212--231},
  year    = {2010},
  doi     = {10.12693/APhysPolA.118.212}
}

@article{Comin2016,
  author  = {Comin, Riccardo and Damascelli, Andrea},
  title   = {Resonant X-Ray Scattering Studies of Charge Order in Cuprates},
  journal = {Annual Review of Condensed Matter Physics},
  volume  = {7},
  pages   = {369--405},
  year    = {2016},
  doi     = {10.1146/annurev-conmatphys-031115-011401}
}

@article{Frano2020,
  author  = {Frano, Alex and Blanco-Canosa, Santiago and
             Keimer, Bernhard and Birgeneau, Robert J.},
  title   = {Charge Ordering in Superconducting Copper Oxides},
  journal = {Journal of Physics: Condensed Matter},
  volume  = {32},
  number  = {37},
  pages   = {374005},
  year    = {2020},
  doi     = {10.1088/1361-648X/ab6140}
}

@article{RevModPhys.76.323,
  title = {Spintronics: Fundamentals and applications},
  author = {\ifmmode \check{Z}\else \v{Z}\fi{}uti\ifmmode \acute{c}\else \'{c}\fi{}, Igor and Fabian, Jaroslav and Das Sarma, S.},
  journal = {Rev. Mod. Phys.},
  volume = {76},
  issue = {2},
  pages = {323--410},
  numpages = {0},
  year = {2004},
  month = {Apr},
  publisher = {American Physical Society},
  doi = {10.1103/RevModPhys.76.323},
  url = {https://link.aps.org/doi/10.1103/RevModPhys.76.323}
}

@article{Ikeda2005,
  author  = {Ikeda, Naoshi and Ohsumi, Hiroyuki and Ohwada, Kenji
             and Ishii, Kenji and Inami, Toshiya and Kakurai, Kazuhisa
             and Murakami, Youichi and Yoshii, Kenji and Mori, Shigeo
             and Horibe, Yoichi and Kit{\^o}, Hijiri},
  title   = {Ferroelectricity from Iron Valence Ordering in the
             Charge-Frustrated System {LuFe2O4}},
  journal = {Nature},
  volume  = {436},
  pages   = {1136--1138},
  year    = {2005},
  doi     = {10.1038/nature04039}
}

\end{document}